\documentclass[aip,jcp,reprint,groupedaddress]{revtex4-1}
\usepackage{graphicx}% Include figure files
\usepackage{dcolumn}% Align table columns on decimal point
\usepackage{bm}% bold math
\usepackage{amsmath,amssymb}

\begin{document}

\title{Multi-resolution polymer Brownian dynamics with hydrodynamic
interactions} 

\author{Edward Rolls}
\email{edward.rolls@pmb.ox.ac.uk}

\author{Radek Erban}
\email{erban@maths.ox.ac.uk}
\homepage{\\ http://people.maths.ox.ac.uk/erban/}

\affiliation{Mathematical Institute, University of Oxford, Radcliffe Observatory Quarter, Woodstock Road, Oxford, OX2 6GG, United Kingdom}

\date{\today}

\begin{abstract} 
A polymer model given in terms of beads, interacting through 
Hookean springs and hydrodynamic forces, is studied.   
Brownian dynamics description of this bead-spring polymer model
is extended to multiple resolutions. Using this multiscale 
approach, a modeller can efficiently look at different 
regions of the polymer in different spatial and temporal resolutions 
with scalings given for the number of beads, statistical segment 
length and bead radius in order to maintain macro-scale properties of 
the polymer filament. The Boltzmann distribution of a Gaussian chain 
for differing statistical segment lengths gives a Langevin equation 
for the multi-resolution model with a mobility tensor for different 
bead sizes. Using the pre-averaging approximation, the translational 
diffusion coefficient is obtained as a function of the inverse of 
a matrix and then in closed form in the long-chain limit. This is 
then confirmed with numerical experiments.
\end{abstract}

\pacs{02, 05, 87}% insert suggested PACS numbers in braces on next line

\maketitle %\maketitle must follow title, authors, abstract and \pacs

\section{Introduction}

\noindent
There have been many studies which use a Brownian dynamics (BD) model for 
polymers with hydrodynamic interactions as a method to coarse grain 
complex interactions at the atomic level to study macromolecules in 
biology and materials science, for example stepping kinetics of 
kinesin~\cite{zhang2012dissecting}, dynamics for $\lambda$-phage
DNA~\cite{jendrejack2002stochastic,schroeder2004effect,quake1997dynamics} 
and for polystyrene~\cite{amelar1991dynamic,hair1989intrinsic}.
BD models describe polymers as beads connected by 
Hookean springs~\cite{Doi1986theory}. In this work, we extend the BD 
modelling framework to allow a polymer molecule to be considered 
on multiple resolutions, so that the statistical segment length and 
bead size vary along the polymer as well as the timestep used for simulating
BD, i.e. we consider multiple resolutions of both spatial and temporal 
scales. 

The use of a multi-resolution model allows us to look at areas of 
interest on the polymer in additional detail, while modelling less 
important areas in less detail and maintaining global 
properties of the polymer. The main benefit of this form of hybrid 
modelling is in computational savings compared to modelling the 
entire domain in high resolution, and has become an increasingly 
popular technique in recent years as a method to look at complex 
models on larger spatial scales with more biologically relevant 
timescales~\cite{erban2014molecular,erban2016atom,korolev2016review,zavadlav2015}. 
This is useful, for example, in modelling the interaction between
a DNA-binding protein and a DNA filament, where only the area near the 
protein needs to be modelled in high resolution and other areas can 
be modelled in a lower resolution. A similar approach to the one taken 
in this paper has been considered in our previous work in the case
where hydrodynamic interactions can be neglected~\cite{rolls2016varying}.

This work acts as an extension to analytic results for properties of 
the single-scale polymer dynamical simulations with hydrodynamic 
interactions. Much of the initial analysis of the single-scale model 
was done by Kirkwood and Riseman~\cite{kirkwood1948intrinsic}, who 
introduced elements of the BD model and gave an approximation to the 
translational diffusion coefficient which uses equilibrium averages 
for internal configurations. Following this, Zimm~\cite{Zimm1959Model} 
found an approximation to the translational diffusion coefficient by 
pre-averaging the inter-particle distances in the hydrodynamic interaction 
tensor. {\"O}ttinger~\cite{ottinger1987translational} found a more 
accurate approximation equivalent to the previous work of 
Fixman~\cite{fixman1981inclusion} by considering the centre of 
hydrodynamic resistance and manipulating the Langevin equation, which 
is the work we will build on to obtain equations for the
diffusion coefficient, partly because its formulation 
has a natural extension to multi-resolution modelling.

In this paper, we start by formulating the model using a statistical 
physics description and using the Boltzmann distribution to form 
a mechanical model, similar to how the single-scale model is built 
by Doi and Edwards~\cite{Doi1986theory}. We also define the 
mobility matrix, which comes from extending the 
Rotne-Prager-Yamakawa tensor~\cite{Rotne1969tensor,Yamakawa1970transport} 
to allow for different bead sizes by considering fluid 
dynamics properties~\cite{zuk2014rotne}. This provides 
a Langevin equation for the new model, as well as formulations 
for the distribution of inter-bead distances, 
including the root mean squared (rms) end-to-end distance. 
From this Langevin description, we form two approximations for 
the translational diffusion coefficient: one as a solution to 
a matrix inversion problem and another in closed form in the 
long chain limit. This form of the diffusion is used to give 
us scaling laws for the statistical segment length, bead radius and 
the number of beads in each region of the polymer. These scaling laws
maintain the global properties of the rms end-to-end distance 
and diffusion.

A number of algorithms have been proposed for efficient BD 
simulations with hydrodynamic interactions for the single-scale 
polymer model in the 
literature~\cite{miao2017iterative,fixman1986construction,ando2012krylov,geyer2009n}.
In Section~\ref{secsimulmeth}, we adapt the Ermak-McCammon 
algorithm~\cite{Ermak1978Brownian} by varying the timestep 
along the polymer in order for BD simulations of the Langevin equation 
to take place. We conclude with illustrative computational results confirming
the presented theory in Section~\ref{seccomresults}.

\section{Multi-resolution bead-spring model}

\noindent
As a model for a polymer, we use an extension to the 
bead-spring model which has existed for over 60 years~\cite{Doi1986theory}. 
This method of modelling has $N$ beads connected with $(N-1)$ Hookean 
springs, neither of which seek to have physical significance as 
such, but represent a coarse-grained description of the direction 
the polymer is coiled. Our multi-resolution extension allows different 
beads to have different sizes and the statistical segment length between 
adjacent beads to vary.

To form the multi-resolution model, we start with a static description 
of the chain as an extension of the Gaussian chain model, where much of 
the analysis follows from the treatment given by Doi 
and Edwards~\cite{Doi1986theory}, but with varying bond lengths. We use this to derive 
a potential for spring constants using the Boltzmann distribution which 
in turn is used to form the dynamic model with hydrodynamic interactions 
given from a multi-resolution extension of the Rotne-Prager-Yamakawa tensor~\cite{Rotne1969tensor,Yamakawa1970transport,zuk2014rotne}.

\subsection{Static Description}

\noindent
We consider a bead-spring polymer model with $N$ beads, where 
the positions of the beads of the chain are given by $\mathbf{r}_n$, 
so that $\mathbf{R}_n = \mathbf{r}_{n+1} - \mathbf{r}_n$, 
for $n=1,2, \dots N-1$, are the corresponding bond vectors.
In multi-resolution bead-spring model, the distribution $\psi_n$ for 
the $n^{\text{th}}$ bond vector 
$\mathbf{R}_n$, $n=1,2,\dots,N-1$, can vary along the chain:
\begin{equation}
\psi_n(\mathbf{R}_n) 
= 
\left(\frac{3}{2 \pi b_n^2}\right)^{3/2} \exp \left(-\frac{3 \mathbf{R}_n^2}{2 b_n^2}\right),
\label{distbondvectors}
\end{equation}
where the statistical segment lengths $\langle \mathbf{R}_n^2 \rangle = b_n^2$ 
are allowed to vary along the filament. In the special case
$b=b_n$ for all $n=1,2,\dots,N-1$, equation (\ref{distbondvectors}) reduces
to the standard Gaussian chain model~\cite{Doi1986theory}.
Using (\ref{distbondvectors}), the distribution of 
$\mathbf{r}_{mn} = \mathbf{r}_n- \mathbf{r}_m$, for $n \ne m$, 
is given by 
\begin{equation}
\Phi(\mathbf{r}_{mn}) = \left(\frac{3}{2 \pi \mu_{mn}^2 }\right)^{3/2} 
\exp \left(-\frac{3 (\mathbf{r}_n-\mathbf{r}_m)^2}{2 \mu_{mn}^2}\right),
\label{distrnm}
\end{equation}
where we define
\begin{equation*}
\mu_{mn}^2 =  \sum_{k=\min\{m,n\}}^{\max\{m,n\}-1} b_k^2.
\end{equation*}
If we define the rms end-to-end distance 
$\mu = \langle \mathbf{R}^2 \rangle^{1/2}$, where $\mathbf{R}$ 
is the vector from the first to the last bead, then we find
\begin{equation}
\label{end2endDist}
\mu 
= 
\sqrt{
\langle (\mathbf{r}_N-\mathbf{r}_1)^2 \rangle
} 
= 
\mu_{1N}
=
\sqrt{\sum_{k=1}^{N-1}b_k^2}.
\end{equation}
Using (\ref{distbondvectors}), the conformational distribution 
function of the chain is given by:
\begin{align}
\label{confDist}
\Psi(\{\mathbf{r}_n\}) &= \prod_{n=1}^{N-1} \psi_n(\mathbf{R}_n) 
\\
&= \left( \frac{3}{2 \pi \prod_{n=1}^{N-1} b_n^2}\right)^{3/2} 
\exp\left(- \sum_{n=1}^{N-1} \frac{3 \mathbf{R}_n^2}{2 b_n^2}\right).
\nonumber 
\end{align}
Considering a polymer chain at equilibrium, a mechanical model with potential
\begin{equation}
\label{potential}
U\left(\{\mathbf{r}_n\}\right) 
= 
\frac{3 k_B T}{2} \sum_{n=1}^{N-1} \frac{(\mathbf{r}_{n+1}-\mathbf{r}_n)^2}{b_n^2},
\end{equation}
has an identical Boltzmann distribution to equation~(\ref{confDist}).

\subsection{Dynamic Model Description}

\noindent
Using potential (\ref{potential}), we form a Langevin equation 
for the dynamic model~\cite{ottinger2012stochastic} for 
$n^{\text{th}}$ bead at time $t$
\begin{align}
\mbox{d} \mathbf{r}_n 
= 
&
\left(\sum_{m=1}^N 
\mathbf{H}_{nm} \, \frac{\partial U}{\partial \mathbf{r}_m} 
+ 
\nabla_m \cdot \mathbf{H}_{nm} \right) \mbox{d}t 
\nonumber
\\
&
+ \sum_{m=1}^N 
\mathbf{B}_{nm} \, \mbox{d}\mathbf{W}_m,
\label{langeq}
\end{align}
where $\mathbf{H}_{mn} \in {\mathbb R}^{3\times 3}$ 
is a positive-definite symmetric mobility matrix, 
$\mbox{d}\mathbf{W}_m \in {\mathbb R}^{3}$ is 
a Wiener process  and we define 
$\mathbf{B}_{mn} \in {\mathbb R}^{3\times 3}$ 
so that
\begin{equation}
\mathbf{H}_{mn} = 
\frac{1}{2 k_B T}
\sum_{k=1}^N 
\mathbf{B}_{mk} \, \mathbf{B}_{nk}^{\mathrm{T}},
\label{Bmndef}
\end{equation}
where $T$ is the absolute temperature, superscript $\mathrm{T}$ 
(roman font) denotes the transpose of a matrix and $k_B$ 
is the Boltzmann constant. The matrix $\mathbf{B}_{mk}$ exists 
due to the positive-definiteness of $\mathbf{H}_{mn}$, and can 
be found by performing a decomposition (e.g. the Cholesky 
decomposition). In choosing the mobility matrix, we would like 
to vary the bead sizes along the filament, which can be used to 
ensure that macroscopic properties of the polymer remain constant. 
Therefore, the bead size $\sigma_n$ becomes a parameter 
of $n^{\text{th}}$ bead.

Rolls et al.\cite{rolls2016varying} use the diagonal mobility tensor 
which works as an extension of the Rouse model: 
\begin{equation}
\mathbf{H}_{mn} = 
\begin{cases}
\frac{1}{6 \pi \eta \sigma_n} \mathbf{I}, 
& \quad\mbox{if } m=n; 
\\
\mathbf{0}, & \quad\mbox{if } m \neq n;
\end{cases}
\label{rollstensor}
\end{equation}
where $\eta$ is the dynamic viscosity and 
$\mathbf{I} \in {\mathbb R}^{3\times 3}$ is the identity matrix. 
One of the purposes of 
this paper is to extend model (\ref{rollstensor}) to include 
hydrodynamic interactions. Many models which include hydrodynamic 
interactions where beads are of equal sizes use 
the Rotne-Prager-Yamakawa\cite{Rotne1969tensor,Yamakawa1970transport} 
tensor. This has been extended by Zuk et al.~\cite{zuk2014rotne} 
to allow for beads of different sizes. To formulate it,
we denote for beads $m$ and $n$: 
\begin{align*}
\sigma_{mn} = &|\sigma_n - \sigma_m|, \\
r_{mn} = & |\mathbf{r}_{n} - \mathbf{r}_{m}|, 
\end{align*}
where $\sigma_m$ (resp. $\sigma_n$) is the radius of
of bead $m$ (resp. $n$) and $r_{mn}$ is the distance
between beads. We also denote by $\mathbf{\widehat{r}}_{mn}$ 
the unit vector between beads, i.e.
$$
\mathbf{\widehat{r}}_{mn} = \displaystyle
\frac{\mathbf{r}_{n} - \mathbf{r}_{m}}{r_{mn}}.
$$
Then the Rotne-Prager-Yamakawa-type mobility 
tensor is given by~\cite{zuk2014rotne}
\begin{widetext}
\begin{equation}
\label{fullMobility}
\mathbf{H}_{mn} = 
\begin{cases}
\frac{1}{6 \pi \eta \sigma_n} \,\mathbf{I},
& \quad\mbox{if } m=n; 
\\
\rule{0pt}{5.1mm}
\frac{1}{8 \pi \eta r_{mn}} 
\left[ \left( 1 + \frac{\sigma_n^2 +\sigma_m^2}{3 r_{mn}^2}\right) 
\mathbf{I} 
+ 
\left( 1 - \frac{\sigma_n^2 +\sigma_m^2}{ r_{mn}^2}\right) 
\mathbf{\widehat{r}}_{mn}\otimes \mathbf{\widehat{r}}_{mn} \right],
& 
\quad\mbox{if } \sigma_n + \sigma_m < r_{mn};
\\
\rule{0pt}{5.1mm}
 \frac{1}{6 \pi \eta \sigma_n \sigma_m} 
\left[ \frac{16 r_{mn}^{3}(\sigma_n + \sigma_m) 
- (\sigma_{mn}^2 + 3 r_{mn}^2)^2}{32 r_{mn}^{3}} 
\,\mathbf{I} 
+ 
\frac{3(\sigma_{mn}^2 - r_{mn}^2)^2}{32 r_{mn}^{3}}
\mathbf{\widehat{r}}_{mn}
\otimes \mathbf{\widehat{r}}_{mn} 
\right],
\qquad
& \quad\mbox{if } \sigma_{mn} < r_{mn} \leq \sigma_n + \sigma_m;
\\ 
\rule{0pt}{5.1mm}
\frac{1}{6 \pi \eta \max\{\sigma_n,\sigma_m\}} \,\mathbf{I},
& \quad\mbox{if } r_{mn} \leq \sigma_{mn};
\end{cases}
\end{equation}
\end{widetext}
which is positive-definite, symmetric and continuous for 
sufficiently small $\sigma_n$ for all $n$.
It is also incompressible, so 
that $\nabla_m \cdot \mathbf{H}_{nm} = 0$, 
which simplifies the Langevin equation (\ref{langeq}) to
\begin{equation} 
\label{langevin}
\mbox{d} \mathbf{r}_n 
= 
\left(\sum_{m=1}^N 
\mathbf{H}_{nm} \, \mathbf{F}_m
\right) \mbox{d}t
+ \sum_{m=1}^N 
\mathbf{B}_{nm} \, \mbox{d}\mathbf{W}_m,
\end{equation}
where $\mathbf{B}_{mn} \in {\mathbb R}^{3\times 3}$ 
are given by (\ref{Bmndef}) which exists by the positive-definite 
symmetric property of $\mathbf{H}_{mn}$ and inter-bead force $\mathbf{F}_m$ 
is found by differentiating the potential~(\ref{potential}) to get
\begin{equation}
\label{force}
\mathbf{F}_m 
= 
\frac{3 k_B T}{b^2_{m-1}} 
\left(\mathbf{r}_{m-1}-\mathbf{r}_{m} \right) 
+ 
\frac{3 k_B T}{b^2_{m}} 
\left(\mathbf{r}_{m+1}-\mathbf{r}_{m}\right).
\end{equation}

\section{Approximation of the Diffusion Coefficient 
and Ideal Parameterisation}

\noindent
We take a similar approach to 
\"{O}ttinger~\cite{ottinger1987translational} to find 
an approximation for the diffusion of the multi-resolution
model. This is then used in conjunction with knowledge 
of the distribution of the rms end-to-end vector to inform 
the scaling for a multiscale simulation, so that properties 
of interest match up to the `ground truth' 
high-resolution model.

\subsection{Diffusion Approximation}

\noindent
To find an approximation to the diffusion for the mobility 
tensor with hydrodynamic terms included, we use the pre-averaging
approximation~\cite{Doi1986theory}, introduced by 
Zimm~\cite{Zimm1959Model}. Considering near-equilibrium dynamics, 
we replace the mobility tensor $\mathbf{H}_{mn}$ with its equilibrium 
average $\langle \mathbf{H}_{mn} \rangle_{\text{eq}}$, using 
$\Psi$ from equation~(\ref{confDist}):
\begin{equation*}
\mathbf{H}_{mn} 
\rightarrow 
\langle \mathbf{H}_{mn} \rangle_{\text{eq}} 
= \int \mathbf{H}_{mn} \Psi(\{\mathbf{r}_{n}\}) \,
\mbox{d} \{\mathbf{r}_{n}\}.
\end{equation*}
If we assume that $\sigma_n + \sigma_{n+1} < b_n$ for 
all $n \leq N-1$, then our equilibrium distribution 
$\langle \mathbf{H}_{mn} \rangle_{\text{eq}}$ for 
off-diagonal entries $m \neq n$ becomes
\begin{align*}
\langle \mathbf{H}_{mn} \rangle_{\text{eq}} 
&
= 
\frac{1}{8 \pi \eta}  
\left[ \left\langle \frac{1}{r_{mn}} \right\rangle_{\!\!\text{eq}}  
\left\langle \mathbf{I} + \mathbf{\widehat{r}}_{mn}
\otimes \mathbf{\widehat{r}}_{mn} \right\rangle_{\text{eq}} 
\right. \nonumber \\
& + \left. (\sigma_n^2 +\sigma_m^2) 
\left\langle \frac{1}{r^3_{mn}} \right\rangle_{\!\!\text{eq}} 
\left\langle \frac{\mathbf{I}}{3} 
- \mathbf{\widehat{r}}_{mn}\otimes \mathbf{\widehat{r}}_{mn} 
\right\rangle_{\!\!\text{eq}}  \right],
\end{align*}
where we have used that the distribution of 
$\mathbf{\widehat{r}}_{mn}$ is independent of $r_{mn}$. 
Using $\langle \mathbf{\widehat{r}}_{mn}
\otimes 
\mathbf{\widehat{r}}_{mn} \rangle_{\text{eq}} = \mathbf{I}/3$, 
the second term cancels and we obtain 
$\langle \mathbf{H}_{mn} \rangle_{\text{eq}} = 
\widehat{H}_{mn} \mathbf{I}$, where
\begin{equation*}
\widehat{H}_{mn}
= 
\begin{cases}
\displaystyle\frac{1}{6 \pi \eta \sigma_n},
& \qquad
\mbox{for } m=n;
\\ 
\rule{0pt}{6mm}
\displaystyle\frac{1}{6 \pi \eta} \left\langle \frac{1}{r_{mn}} 
\right\rangle_{\!\!\text{eq}},
& \qquad
\mbox{for } m \neq n.
\end{cases}
\end{equation*}
Using (\ref{distrnm}), we obtain
\begin{equation}
\label{mobilityMatrix}
\widehat{H}_{mn} = 
\begin{cases}
\displaystyle\frac{1}{6 \pi \eta \sigma_n}, 
& \qquad
\mbox{for } m=n;
\\
\rule{0pt}{6mm}
\displaystyle
\frac{1}{\mu_{mn} \, \eta \, \pi \sqrt{6 \pi}},
& \qquad
\mbox{for } m \neq n.
\end{cases}
\end{equation}
In the single-scale model where $b_n=b$ and $\sigma_n=\sigma$ for all $n$,
equation (\ref{mobilityMatrix}) generalises to the equation for the 
pre-averaged tensor in Doi and Edwards~\cite{Doi1986theory,Zimm1959Model}.
Consequently, by pre-averaging equation~(\ref{langevin}), 
we find
\begin{equation} 
\label{langevinApprox}
\mbox{d} \mathbf{r}_n 
= 
\left(\sum_{m=1}^N 
\widehat{H}_{nm} \, \mathbf{F}_m
\right) \mbox{d}t 
+ \sum_{m=1}^N 
\widehat{B}_{nm} \, \mbox{d}\mathbf{W}_m,
\end{equation}
where 
\begin{equation}
\widehat{H}_{mn} 
= 
\frac{1}{2 k_B T}
\sum_{k=1}^N \widehat{B}_{mk} \widehat{B}_{nk}.
\label{HmnBmnrelation}
\end{equation}
Following \"{O}ttinger~\cite{ottinger2012stochastic}, we define 
the hydrodynamic center of resistance $\mathbf{r}_h$ by
\begin{equation*}
\mathbf{r}_h = \sum_{n=1}^{N} l_n \mathbf{r}_n,
\qquad
\mbox{where}
\qquad
l_n 
= 
\frac{\sum_{m=1}^N \widehat{H}^{-1}_{nm}}{
\sum_{m,k=1}^N \widehat{H}^{-1}_{km}}.
\end{equation*}
Multiplying equation~(\ref{langevinApprox}) through by $l_n$ 
and summing over all $n$, we get
\begin{equation*} 
\mbox{d} \mathbf{r}_h 
\!=\! 
\left(\sum_{m,n=1}^N 
l_n \widehat{H}_{nm} \, \mathbf{F}_m
\!\right) \mbox{d}t 
+ \sum_{m=1}^N 
\left(
\sum_{n=1}^N 
l_n
\widehat{B}_{nm} 
\!\right)
\mbox{d}\mathbf{W}_m.
\end{equation*}
Using (\ref{force}), the first term on the right hand side is zero and 
the second term is a linear combination of Wiener processes, which is
itself a Wiener process with translational diffusion coefficient
\begin{equation*}
D_h =
\frac{1}{2}
\sum_{m=1}^N
\left(
\sum_{n=1}^N 
l_n \widehat{B}_{nm} 
\right)^2.
\end{equation*}
Using (\ref{HmnBmnrelation}) and the definition of $l_n$, we obtain
\begin{equation}
D_h = 
 k_B T \left(\sum_{m,n=1}^N \widehat{H}^{-1}_{mn} \right)^{\!\!-1}.
\label{diffusionApprox}
\end{equation}
This forms a matrix equation to provide the pre-averaged 
approximation for the translational diffusion coefficient.

\subsection{Behaviour in the Long Chain Limit}

\noindent
Our analysis in the previous section has used a general multi-resolution
model consisting of $N$ beads with sizes $\sigma_n$, $n=1,2,\dots,N$,
connected by $N-1$ springs with statistical segment lengths $b_n$,
$n=1,2,\dots,N-1$. In applications to multiscale computations, we are 
mostly interested in chains which are split into $M$ regions 
(where $M \ll N$) of constant statistical segment length. 
In what follows, we will use lower case greek subscripts 
$\alpha$ (resp. $\beta$ and $\gamma$) to denote regions, while 
$n$ (resp. $m$ and $k$) are indices refering to numbers of 
individual beads and springs along the polymer chain. We assume 
that the $\alpha$-th region contains $N_\alpha$ springs, with 
statistical segment length $b_\alpha$, $\alpha=1,2,\dots,M$. 
Summing over all regions, we have
$$
\sum_{\alpha=1}^{M}N_\alpha = N-1.
$$
In this section, we simplify equation (\ref{diffusionApprox})
in the long chain limit, $N \rightarrow \infty$, which is taken
in such a way that the fraction of springs in each region, $N_\alpha/(N-1)$, 
remains a constant, i.e. if $(N-1)$ doubles in size then each individual 
region also doubles in size.

Equation~(\ref{mobilityMatrix}) defines a function of two integer
variables $m$ and $n$. We will map the discrete function 
$\widehat{H}_{mn}$ into a continuous function $H(x,y)$ by generalizing
the approach of \"{O}ttinger~\cite{ottinger2012stochastic} and
Fixman~\cite{fixman1981inclusion}. Assuming that bead $m$ lies in 
region $\alpha$, its continuous approximation in interval
$[-1,1]$ will be defined by
\begin{equation}
\frac{2b_\alpha^2}{\mu^2} \, m - 1 
+ 
\frac{2}{\mu^2} \sum_{\gamma=1}^{\alpha-1}N_\gamma (b_\gamma^2-b_\alpha^2)
\;
\longrightarrow
\;
x,
\label{xtransf}
\end{equation}
where 
$
\mu^2 
= 
\sum_{\gamma=1}^{M} N_\gamma b_\gamma^2.
$
The continuous analogue of a summation of
arbitrary function $f_n$ over all beads 
will then be a weighted integral
\begin{equation}
\label{intMapping}
\sum_{n=1}^{N}f_n 
\;
\longrightarrow 
\;
\int_{-1}^{1} f(x) \, b(x) \,\mbox{d}x, 
\end{equation}
where we define $b(x)$ as a piecewise constant function
given by $b(x) = \mu^2 b_\alpha^{-2} /2$ in interval
\begin{equation*}
x \in \left( \frac{2}{\mu^2} 
\sum_{\gamma=1}^{\alpha-1} N_\gamma b_\gamma^2 - 1, 
\frac{2}{\mu^2} \sum_{\gamma=1}^{\alpha} N_\gamma b_\gamma^2 
- 1 \right].
\end{equation*}
In addition to (\ref{xtransf}), we also write
for bead $n$ in region $\beta$
\begin{equation*}
\frac{2b_\beta^2}{\mu^2} \, n - 1 + \frac{2}{\mu^2} \sum_{\gamma=1}^{\beta-1}N_\gamma(b_\gamma^2-b_\beta^2)
\;
\longrightarrow
\;
y.
\end{equation*}
This leeds to a transformation $(m,n) \rightarrow (x,y)$, which
gives the continuous approximation of $\widehat{H}_{mn}$ 
in $[-1,1] \times [-1,1]$ as
\begin{equation}
H(x,y) = \sqrt{\frac{2}{\mu^2}} \frac{1}{\eta \pi \sqrt{6 \pi |x-y|}}.
\label{Hxy}
\end{equation}
The definition of the inverse of the matrix $\widehat{H}_{mn}$
(given as $\sum_{k=1}^{N} \widehat{H}_{mk} \widehat{H}_{kn}^{-1} = \delta_{mn}$),
is rewritten in continous variables as
\begin{equation*} 
\int_{-1}^{1}  
H(x,z) \, H^{-1}(z,y) \, b(z)\, \mbox{d}z = \frac{\delta(x-y)}{b(x)}.
\end{equation*}
Multiplying both sides by $b(y)$, integrating over $y$ and 
using (\ref{Hxy}), we obtain
\begin{equation}
\int_{-1}^{1} \frac{\phi(z)}{\sqrt{|x-z|}} \, \mbox{d}z 
= 
\eta \, \mu \, \pi \, \sqrt{3 \pi},
\label{integralEqn}
\end{equation}
where 
$$
\phi(z) = \int_{-1}^{1} b(y) \, b(z) \, H^{-1}(z,y) \, \mbox{d}y.
$$
Using the the method of Auer 
and Gardner~\cite{auer1955note,fixman1981inclusion}, we 
solve equation (\ref{integralEqn}) for $\phi(z)$ to obtain
\begin{equation*}
\phi(z) = \frac{\eta \, \mu \, \sqrt{3 \pi}}{\sqrt{2} \, (1-z^2)^{1/4}}.
\end{equation*}
To return to the quantity of interest, 
$\sum \sum \widehat{H}_{nm}^{-1}$, we apply the mapping from
equation~(\ref{intMapping}) to give
\begin{align*}
\sum_{n=1}^{N} \sum_{m=1}^{N} \widehat{H}_{nm}^{-1}
&= \int_{-1}^{1} \int_{-1}^{1} b(x)b(y)H^{-1}(x,y)\, \mbox{d}y \, \mbox{d}x \\
&= \int_{-1}^{1} \phi(x) \, \mbox{d}x 
= \frac{\eta \, \mu \, 4 \pi^2 \sqrt{3}}{\Gamma^2(1/4)},
\end{align*}
where $\Gamma$ is the gamma function. 
Substituting in (\ref{diffusionApprox}), we obtain
diffusion constant in the long chain limit
\begin{equation}
\label{longChainDiffusion}
D_h = 
\frac{\Gamma^2(1/4)}{4 \pi^2 \sqrt{3}} \,
\frac{k_B T}{\eta \, \mu} 
\;
\approx 
\;
0.1922 \, \frac{k_BT}{\eta \, \mu}.
\end{equation}

\subsection{Scaling of Parameters}
\label{paramScaling}

\noindent
As the use of a bead-spring model is to give a 
coarse-grained representation of a filament, 
statistical segment length and bead radius are 
not physical qualities so we allow these parameters 
of the model to vary in order to achieve desired 
statistics of interest for the polymer. In our previous 
work~\cite{rolls2016varying}, the whole-system statistics 
of interest have been the rms end-to-end 
distance $\mu$ and translational diffusion coefficient 
$D_h$ of a polymer chain. In this paper, we will consider
three quantities which multi-resolutions simulations should 
preserve: the rms end-to-end distance $\mu$, 
diffusion coefficient $D_h$ and the strength 
of hydrodynamic interactions~\cite{larson2005rheology}, 
defined in terms of the parameter $h^*$ by
\begin{equation*}
h^* = \sqrt{\frac{3}{\pi}} \frac{\sigma}{b},
\end{equation*}
where $b$ is the statistical segment length and $\sigma$ is the bead radius.
From a theoretical standpoint, a value for $h^* \approx 0.25$ minimises 
the effect of chain length~\cite{ottinger1987generalized}.
Similar values can also match experimental results for viscoelastic 
properties, for example the Flory-Fox parameter can match 
experimental values~\cite{larson2005rheology} for $h^* \approx 0.267$. 
In the multi-resolution model, in order to maintain a consistent 
value for the strength of hydrodynamic interactions, we therefore 
scale parameters in order to keep $h^*$ constant throughout simulations.

The multi-resolution polymer simulations will be compared
to the `ground-truth' model, which will be the single-scale 
model of the polymer in the maximum detail required. 
In single-scale models, we can modify the whole-system statistics 
by varying the statistical segment length $b$, the bead radius 
$\sigma$ and the total number of beads $N$. In 
single-scale models, we need to select a level of detail 
for the entire chain as an additional constraint, but 
by modelling on multiple scales, we instead get to 
select the resolution of different regions of the polymer, 
so that only regions of particular interest need to be 
in the highest level of detail. To parameterize
the `ground-truth' model, we select $b$ to give the desired 
value for the rms end-to-end distance $\mu$, from 
equation~(\ref{end2endDist}), i.e. $b = \mu (N-1)^{-1/2}.$ 
Selecting a value for $\sigma$ is a bit more subtle than 
for models without hydrodynamic interactions\cite{rolls2016varying}, 
as the inclusion of the hydrodynamic interactions mean 
the leading order long chain diffusion approximation 
in equation~(\ref{longChainDiffusion}) is independent 
of $\sigma$, i.e. we cannot use $D_h$. We use the strength 
of hydrodynamic interaction $h^*$ to select an appropriate
value of $\sigma$. For all simulations in this paper, 
we use $\sigma=b/4$.

Once we have defined the `ground-truth' model, with statistical 
segment length $b$, bead radius $\sigma$ and total beads $N$, 
we can seek to define the scalings for the multi-resolution 
model, where different regions coarse-grain the original model 
to differing extents. We divide the polymer into $M$ regions
and assume that the `ground-truth' chain contains 
$\widetilde{N}_\alpha$ consecutive springs in the $\alpha^{\mathrm{th}}$
region, for $\alpha=1,2,\dots,M$. Each region of the multi-resolution
model has an associated (integer-valued) resolution $s_\alpha$ such that 
$s_\alpha^2 | \widetilde{N}_\alpha$. Larger values of $s_\alpha$ 
represent coarser regions, and $s_\alpha=1$ gives the 
`ground-truth' model. In the  $\alpha^{\text{th}}$ region
of the multi-resolution model, we have $N_\alpha$ springs with 
statistical segment length $b_\alpha$ given by
\begin{equation}
\label{scalings}
N_\alpha = \frac{\widetilde{N}_\alpha}{s_\alpha^2}, \qquad
b_\alpha = s_\alpha b, \qquad 
\sigma_\alpha = s_\alpha \sigma,
\end{equation}
where the definition of the bead radius, $\sigma_\alpha$, is slightly 
modified for the boundary beads; scalings~(\ref{scalings}) apply 
to beads where both adjacent springs 
have the same statistical segment length $b_\alpha$. 
On the boundaries between regions $\alpha$ and $\alpha+1$, 
for $\alpha=1,2,\dots,M-1$, we take the bead radius to be 
$((\sigma_\alpha^2 + \sigma_{\alpha+1}^2)/2)^{1/2}$, and 
for end beads at the start and end of the polymer we take 
$\sigma_1/\sqrt{2}$ and $\sigma_M/\sqrt{2}$, respectively.
By applying scalings (\ref{scalings}), equation~(\ref{end2endDist})
gives the expected rms end-to-end distance for the filament 
at equilibrium to be $\mu = b (N-1)^{1/2}$, i.e. it is equal 
to the `ground-truth' model. The translational diffusion coefficient for 
the polymer in the long chain limit, equation (\ref{longChainDiffusion}),
is also invariant to the number of regions, as well as the 
size and resolution of each region, and the strength
of hydrodynamic interactions is constant along the filament.

\section{Simulation Method}
\label{secsimulmeth}

\noindent
We solve the Langevin equation for the polymer in 
equation~(\ref{langevin}) by using a modified version of 
the Ermak-McCammon algorithm~\cite{Ermak1978Brownian}, for 
which different regions have different timestep sizes. 
The key idea for the modified algorithm is to keep track 
of the behavior of beads modelled with a higher resolution 
(and with smaller timesteps) to give an average of the 
hydrodynamic forces exerted on the coarsely modelled beads 
between the larger timesteps on which they are modelled.

If the `ground-truth' model uses timestep $\Delta t$, then 
with the notation of Section~\ref{paramScaling} we 
define the timestep associated with the $\alpha^\mathrm{th}$
region as
\begin{equation*}
\Delta t_\alpha = s_\alpha^3 \Delta t,
\end{equation*}
where $s_\alpha$ is the (integer-valued) resolution of 
the $\alpha^\mathrm{th}$ region, $\alpha=1,2,\dots,M$. 
A requirement for the resolution value is that for any 
two regions $\alpha_1$ and $\alpha_2$ that either 
$s_{\alpha_1}^3 | s_{\alpha_2}^3$ or 
$s_{\alpha_2}^3 | s_{\alpha_1}^3$, to ensure the timesteps 
of the coarser regions match up to those for the finer regions.
We choose this scaling to ensure numerical stability of 
simulations so that the size of the tension term for 
a bead is much smaller than the statistical segment 
length with adjacent beads. In the case of a bead lying 
between two regions, we take the timestep to be the 
minimum value of the timesteps given by each region.

The `ground-truth' model updates time at integer multiples
of $\Delta t$, i.e. we compute the polymer state at 
times $t=i \Delta t$, where $i=0,1,2,3,\dots.$ Considering 
the multi-resolution model, we can formally write the 
update rule (from time $i \Delta t$ to time 
$(i+1) \Delta t$) for the $n^{\text{th}}$ bead, 
for $n=1,2,\dots,N$, in region $\alpha_n$ as
\begin{align}
&\mathbf{r}_n((i+1) \Delta t) 
=
\;\mathbf{r}_n(i \Delta t) 
+ Q(s_{\alpha_n}^3,i+1)
\bigg(
\bm{\widetilde{\rho}}_n(i \Delta t)
\nonumber
\\
&
\qquad\qquad
+
\sum_{m=1}^N Q(s_{\alpha_m}^3,i+1)
\,
\mathbf{H}_{nm} 
\mathbf{\widetilde{F}}_{mn}(i \Delta t) 
\bigg),
\label{updaterule}
\end{align}
where $\mathbf{H}_{nm}$ is the mobility tensor given 
in equation~(\ref{fullMobility}), $\mathbf{\widetilde{F}}_{mn}$
and $\bm{\widetilde{\rho}}_n$ are discretized force and
noise terms given below,  and 
the function $Q$ is defined for integers $i$ and $j$ by
\begin{equation*}
Q(j,i) = 
\begin{cases}
 1, & \quad\text{if } \;j\,|\,i, \\
 0, & \quad\text{if } \;j\nmid i. \\
\end{cases}
\end{equation*}
To define discretized force and noise terms, we denote by 
$\alpha_n$ (resp. $\beta_n$) the resolution region for 
the $n^{\text{th}}$ bead (resp. spring). Note that for 
beads with both adjacent springs in the same region we 
will see $\alpha_n=\beta_n$, however between regions a 
bead takes the smaller timestep of the adjacent regions, 
so we may see $\alpha_n \neq \beta_n$. In the update 
rule (\ref{updaterule}),
the timestep is incorporated in the force term. We define
the force multiplied by the time step for the $m$-th bead by 
\begin{align*}
\mathbf{\overline{F}}_m(i \Delta t) 
&= 
\frac{3 k_B T}{b^2_{m-1}} 
\left(\mathbf{r}_{m-1}(i \Delta t)-\mathbf{r}_{m}(i \Delta t) 
\right)
\Delta t_{\beta_{m-1}} 
\\
&+ \frac{3 k_B T}{b^2_{m}} 
\left(\mathbf{r}_{m+1}(i \Delta t)-\mathbf{r}_{m}(i \Delta t)\right)
\Delta t_{\beta_{m}}.
\end{align*}
This force term is used as a part of a tension term which 
includes a memory component for larger timesteps, as explained 
in Table~\ref{tableone}:
\begin{equation*}
\mathbf{\widetilde{F}}_{mn}(i \Delta t) = 
\begin{cases}
\mathbf{\overline{F}}_m(i \Delta t), & \text{if } s_{\alpha_n}<s_{\alpha_m}, 
\\
\displaystyle
\sum_{p=0}^{(s_{\alpha_n}/s_{\alpha_m})^3-1}
\!\!\!\!\!\!\! 
\mathbf{\overline{F}}_m((i-p)\Delta t), & \text{otherwise.}
\end{cases}
\end{equation*}
\begin{table}
 \begin{tabular}{m{0.7cm} m{0.7cm} m{0.7cm} m{0.7cm} m{0.7cm} m{0.7cm} m{0.7cm} m{0.7cm} m{0.7cm} m{0.7cm}}
   & 0 & $\Delta t_1$ & $2\Delta t_1$ & $3\Delta t_1$ & $4\Delta t_1$ & $5\Delta t_1$ & $6\Delta t_1$ & $7\Delta t_1$ & $8\Delta t_1$ \\ 
   $s_1$ & $\bullet$ & $\bullet$ & $\bullet$ & $\bullet$ & $\bullet$ & $\bullet$ & $\bullet$ & $\bullet$ & $\bullet$\\ 
   $s_2$ & $\circ$ & & & & & & & & $\circ$\\ 
  \end{tabular}
 \caption{\label{tableone} 
 {\it An explanation of the model running on multiple timesteps 
 with region $s_1$ in twice finer resolution than $s_2$, giving 
 timesteps $8$-times smaller. We run a timestep using 
 equation~$(\ref{updaterule})$ for each bead, so that 
 for every timestep we simulate the beads in $s_1$ and every 
 $8$ timesteps for $s_2$. The force terms 
 $\overline{\bm{F}}_{m}(i\Delta t)$ where bead $m$ is in $s_1$ 
 is summed over the small timesteps to give 
 $\widetilde{\bm{F}}_{mn}(i\Delta t)$ on the larger timesteps. 
 We use the same concept to find noise terms 
 $\widetilde{X}_{mn}(i\Delta t)$.}}
\end{table}%
The random displacement term $\bm{\widetilde{\rho}}_n$ 
has a multivariate Gaussian distribution defined by the moments
\begin{align*}
\langle \bm{\widetilde{\rho}}_n \rangle &= 0 \\
\langle \bm{\widetilde{\rho}}_m 
\otimes \bm{\widetilde{\rho}}_n \rangle &= 
2 k_B T \, \mathbf{H}_{mn} \max (\Delta t_{\alpha_m},
\Delta t_{\alpha_n}),
\end{align*}
where we use the maximum as this term is only expressed 
on the larger of the two timesteps associated with the beads, 
as laid out below. To calculate the $\bm{\widetilde{\rho}}_n$ 
terms, we use an adapted version of the Ermak-McCammon
algorithm~\cite{Ermak1978Brownian}, so that when we flatten 
the tensor $\mathbf{H}_{mn}$ into a matrix $\mathbf{H}$, we 
do so by re-ordering the beads so that beads with smaller 
timesteps have a smaller index than beads with larger timesteps.
Having made this adjustment, we can reduce the computational 
load of the Cholesky decomposition done by the Ermak-McCammon 
algorithm (an $\mathcal{O}(N^3)$ 
calculation~\cite{saadat2014computationally}), by only 
calculating the submatrix made up of the rows and columns of 
the matrix corresponding to beads which are being 
calculated on that particular timestep. Therefore, if there 
are $N_0$ beads which move on a given timestep, then this 
gives a $3N_0 \times 3N_0$ submatrix.

We use the Cholesky decomposition outlined by Ermak and 
McCammon~\cite{Ermak1978Brownian} to get the lower triagonal matrix 
$\bm{B}$, such that $\mathbf{H} = \mathbf{B B^T}$ with elements given 
by, for $n=1,2,\cdots,3N_0,$
\begin{align*}
B_{nn} &= \left( H_{nn} - \sum_{k=1}^{n-1} B_{nk}^2 \right) ^{1/2}, \\
B_{mn} &= \frac{\left( H_{mn} - \sum_{k=1}^{n} B_{mk}B_{nk} \right)}{B_{mm}},
\end{align*}
to give noise terms $\bm{\widetilde{\rho}}_n$ with the calculation
\begin{equation*}
\widetilde{\rho}_n (i \Delta t) 
= \sum_{m=1}^{n} \nu_{nm} \widetilde{X}_{mn}(i \Delta t),
\end{equation*}
which are then reordered and formatted to give 
$\bm{\widetilde{\rho}}_n$ in a $N_0\times3$ matrix. The random 
terms $\tilde{X}_{mn}$ now include 
a `memory' similar to the tension terms so that
\begin{equation*}
\widetilde{X}_{mn}(i \Delta t) = 
\begin{cases}
 X_{m}(i \Delta t),
  & \!\!\! \text{if } s_{\alpha_n}<s_{\alpha_m}, 
\\
\displaystyle
\sum_{p=0}^{(s_{\alpha_n}/s_{\alpha_m})^3-1}
\!\!\!\!\!\!\! 
  X_m((t-p)\Delta t), & \text{otherwise,}
\end{cases}
\end{equation*}
for the terms $X_m(t)$ drawn from a Gaussian normal 
distribution such that $\langle X_m(t) \rangle = 0$ 
and $\langle X_m(t_1) X_n(t_2) \rangle 
= 2 k_B T \delta_{mn} \Delta t_{\alpha_n} \delta(t_1 - t_2).$

\section{Simulations}
\label{seccomresults}

\noindent
We compare the full BD modelling with the dynamics 
using the pre-averaged tensor 
$\langle \mathbf{H}_{mn} \rangle_{\text{eq}}$ from 
equation~(\ref{mobilityMatrix}) in place of $\mathbf{H}_{mn}$ 
in the Ermack-McCammon algorithm given above.
Similar to previous papers simulating bead-spring 
models~\cite{liu2003translational}, we choose to simulate 
with unit parameters, which in our case has $k_B T=1$ 
and $\eta=1$. We shall also hold the rms end-to-end 
distance constant with $\mu=1$, which we shall maintain 
by varying the statistical segment length $b$ as a function of the bead 
number $N$ as appropriate given the scalings explained 
in Section~\ref{paramScaling}. To ensure numerical stability of 
simulations, we found $\Delta t = 10^{-2} b^3 \eta / k_B T$ 
to be a good value to use.

In order to study the translational diffusion coefficient 
in simulations, we need to extend its definition for the 
simulations of the multi-resolution model. We define 
the mass of the polymer $\Omega$ as
\begin{equation*}
\Omega = \sum_{n=1}^N \sigma_n^2,
\end{equation*}
where $\sigma_n$ is the radius of the $n^{\text{th}}$ bead,
$n=1,2,\dots,N$. Using scalings (\ref{scalings}), we observe
that $\Omega$ is invariant to our choice of resolutions
in the multi-resolution scheme. This allows us to define 
the centre of mass of a polymer $\mathbf{r}_G$ at time $t$ as
\begin{equation*}
\mathbf{r}_G(t) 
= 
\frac{\sum_{n=1}^{N} \mathbf{r}_n(t) \sigma_n^2}{\Omega},
\end{equation*}
where $\mathbf{r}_n(t)$ is the position of the 
$n^{\text{th}}$ bead at time $t$. With this we retain the 
definition of the translational diffusion coefficient as
\begin{equation}
D_G = \lim_{t \rightarrow \infty} \frac{1}{6t} 
\langle (\mathbf{r}_G(t) - \mathbf{r}_G(0))^2\rangle.
\end{equation}
Note that for the single scale simulation this reduces to 
the standard definition for the translational diffusion coefficient.

In this section we compare the translational diffusion coefficient 
approximations given in the form of an inverse matrix in 
equation~(\ref{diffusionApprox}) as well as the long chain 
limit approximation in equation~(\ref{longChainDiffusion}) 
to BD simulations for both the full mobility matrix as 
well as the pre-averaged approximation, as well as the 
rms end-to-end distance, which has expected value $1$ in 
all simulations by design. We run the BD simulations 
using both a pre-averaged and a non-pre-averaged mobility 
tensor for a total of $10^4$ timesteps (in the case of 
multi-resolution simulations, this refers to timesteps 
associated with the higher resolution beads), where each 
result is given as an average over $500$ runs, and 
contrast this to the diffusion approximations.

In the results tables we include 95\% confidence intervals 
for the translational diffusion coefficient and the rms 
end-to-end distances. For the end-to-end distance we calculate 
the 95\% confidence interval for $\langle \mathbf{R}^2 \rangle$ 
and take the square root for both lower and upper bounds to 
give a range (note that this is not symmetric about the rms value).

\begin{table*}
 \begin{tabular}{| c || c || c | c | c || c | c | c |}
   \hline
  $N$ & $D_{\text{MF}}$
  & $D_{\text{PA}}$ & $\mu_{\text{PA}}$ & $CI_{\mu_{\text{NPA}}}$ & $D_{\text{NPA}}$  & $\mu_{\text{NPA}}$ & $CI_{\mu_{\text{NPA}}}$ \\ \hline
  5 &  0.178 & $0.187\pm0.013$ & 0.99 & $[0.96,1.01]$ & $0.180\pm0.013$ & 1.00 & $[0.96,1.03]$ \\ \hline
  10 &  0.184 & $0.195\pm0.015$ & 0.99 & $[0.96,1.03]$ & $0.185 \pm 0.014$ & 1.00 & $[0.97,1.04]$\\ \hline
  30 &  0.189 & $0.192\pm0.013$ & 1.00 & $[0.97,1.04]$ & $0.186 \pm 0.013$ & 1.00 & $[0.96,1.03]$ \\ \hline
  50 &  0.191 & $0.202\pm0.014$ & 1.01 & $[0.98,1.05]$ & $0.180 \pm 0.013$ & 1.01 & $[0.97,1.04]$ \\ \hline
  100 & 0.191 & $0.188\pm0.013$ & 1.01 & $[0.98,1.05]$ & $0.174 \pm 0.012$ & 1.01 & $[0.98,1.05]$\\ \hline
  200 & 0.192 & $0.184\pm0.013$ & 0.99 & $[0.96,1.03]$ & $0.193 \pm 0.014$  & 1.00 & $[0.96,1.04]$\\ \hline
 \end{tabular}
 \caption{\label{tabletwo} 
 {\it Results for diffusion and the rms end-to-end distance 
  in single-scale simulations. The subscript 
  {\rm $\text{MF}$} represents the matrix formulation, 
  {\rm $\text{PA}$} is for pre-averaged 
  and {\rm $\text{NPA}$} is non-pre-averaged. The $95$\% confidence 
  intervals for $\mu_{\text{{\rm PA}}}$ and $\mu_{\text{{\rm NPA}}}$ 
  are given by $CI_{\mu_{\text{{\rm PA}}}}$ and $CI_{\mu_{\text{{\rm NPA}}}}$, 
  respectively.}}
 \end{table*}
  
\begin{figure}
\includegraphics[width=0.48332\textwidth]{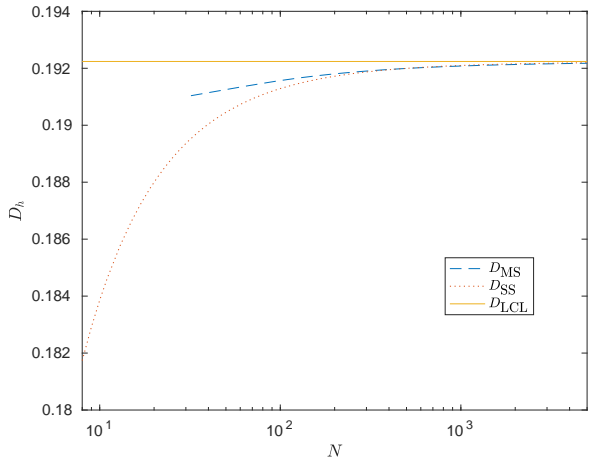}
\caption{\label{fig1} {\it The difference between the matrix 
formulation for the diffusion coefficient $(\ref{diffusionApprox})$
and the long chain limit $(\ref{longChainDiffusion})$ 
as $N$ gets large, for both the single scale-system 
(red dotted line) and the multiscale simulation, which has
the middle $25$\% of the polymer in high resolution
(blue dashed line). Parameters are given in Section~{\rm \ref{seccomresults}}. 
Note that the multi-scale system needs $N \geq 32$ by construction 
for the coarse grained particles to be placed.}}
\end{figure}

We consider three illustrative examples. The first one 
is a single-scale system, so that $M=1$ and $s_1=1$. We use it
as a control to compare the other simulations to. Our results 
are presented in Figure~\ref{fig1} and Table~\ref{tabletwo}. As 
we can see in the table, the matrix formulation $D_{\text{MF}}$ 
is covered within the 95\% confidence interval all except one 
of the pre-averaged and non-pre-averaged values of $N$. The 
analytic value of $\mu=1$ fits in the confidence interval 
for all simulations.

The next system to consider is one in which the middle 
25\% of the polymer is in high resolution, while the 
remainder is coarse-grained by a factor of 2. Therefore 
we define $M=3$ with $s_1=2,$ $s_2=1,$ $s_3=2$, 
and $\widetilde{N}_1=3N/8$, $\widetilde{N}_2=N/4$, 
$\widetilde{N}_3=3N/8$.
The diffusion and the rms end-to-end distance of this 
polymer is given in Table~\ref{tablethree}. The matrix 
formulation for the diffusion $D_{\text{MF}}$ is contained 
in the confidence intervals for all values of $N$, and the 
value of $\mu=1$ is contained in the confidence interval 
for all values of $N$. The convergence
of matrix formulation to the long chain limit is shown
in Figure~\ref{fig1}. 

The final system considered had an 8-times resolution increase 
in the middle 10\% of the the polymer. This uses $M=3$ with 
$s_1=8,$ $s_2=1,$ $s_3=8$, and $\widetilde{N}_1=9N/20,$
$\widetilde{N}_2=N/10,$ $\widetilde{N}_3=9N/20$.
We perform the simulation for $N=1280$, from which we can 
report $D = 0.189$, $\mu = 1.04$ for the pre-averaged case 
and $D = 0.180$, $\mu = 1.01$ where we do not 
use pre-averaging. The matrix formulation for the translational 
diffusion coefficient gives a value $D_{\text{MF}}=0.190$. 
The end-to-end distance in the pre-averaged case narrowly 
falls out of the 95\% confidence interval, but the other 
three stastics lie within this range.

\begin{figure}
\includegraphics[width=0.48332\textwidth]{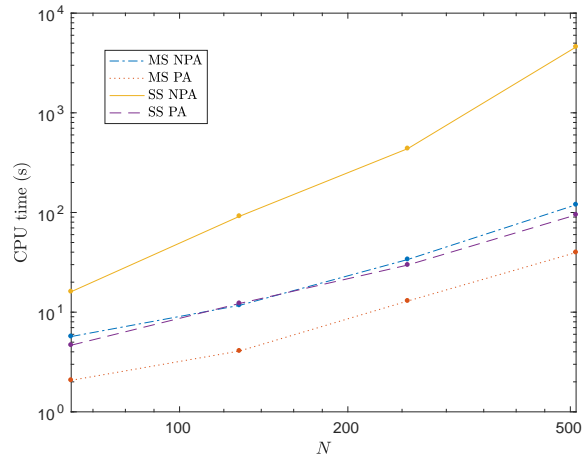}
\caption{{\it The {\rm CPU} times to simulate identical systems using 
four different algorithms: 
the multi-scale model using pre-averaging (red dotted line),
the single-scale model with pre-averaging (purple dashed line), 
the multi-scale model without pre-averaging (blue dot-dashed line)
and the single-scale model without 
pre-averaging (yellow solid line)}.
\label{fig2}}
\end{figure}

As can be seen from the simulations, there is good 
agreement both between the pre-averaged and non pre-averaged 
tensors, as well as between the diffusion approximations and 
the results from the simulations. In total out of $44$ observations, 
we had two fall outside of the 95\% confidence interval.
The overall goal of doing this coarse-graining is to 
improve the speed of simulations. In Figure~\ref{fig2}, 
we compared the timings between the single-scale and 
multi-scale models for identical parameters as were used 
to produce Table~\ref{tablethree}. There is a pronounced 
difference between the multi-scale model and single-scale 
model without pre-averaging, most of which comes from 
having to use the Cholesky decomposition on smaller matrices, 
while the smaller difference in the model with pre-averaging 
comes from updating fewer beads in each timestep.

\begin{table*}
 \begin{tabular}{| c || c || c | c | c || c | c | c |}
   \hline
  $N$ & $D_{\text{MF}}$
  & $D_{\text{PA}}$ & $\mu_{\text{PA}}$ & $CI_{\mu_{\text{NPA}}}$ & $D_{\text{NPA}}$  & $\mu_{\text{NPA}}$ & $CI_{\mu_{\text{NPA}}}$ \\ \hline
  32 &  0.191 & $0.183\pm0.013$ & 0.99 & $[0.95,1.02]$ & $0.189\pm0.013$ & 1.02 & $[0.98,1.05]$ \\ \hline
  64 &  0.191 & $0.195\pm0.013$ & 0.99 & $[0.95,1.02]$ & $0.186 \pm 0.014$ & 0.99 & $[0.96,1.03]$\\ \hline
  128 &  0.192 & $0.185\pm0.014$ & 1.00 & $[0.96,1.04]$ & $0.192 \pm 0.013$ & 1.00 & $[0.97,1.04]$ \\ \hline
  256 &  0.192 & $0.186\pm0.013$ & 1.00 & $[0.96,1.03]$ & $0.184 \pm 0.014$ & 0.98 & $[0.94,1.01]$ \\ \hline
 \end{tabular}
 \caption{\label{tablethree} 
 {\it Results for diffusion and the rms end-to-end distance
 in multi-resolution simulations, which have the middle $25$\% 
 of the polymer in high resolution. 
 Subscripts are the same as in Table~{\rm \ref{tabletwo}}. Simulations 
 run for $10^4$ timesteps for beads at the highest resolution.}}
 \end{table*}

\section{Discussion}

\noindent
In this paper we have extended the bead-spring model for 
a polymer including hydrodynamic interactions to a multi-resolution 
model in order to gain computational efficiency for BD modelling.
By considering a multi-resolution Gaussian chain model, we have
utilised the Boltzmann distribution in order to form a Langevin 
equation for the multi-resolution model. From this we used a 
similar approach to {\"O}ttinger~\cite{ottinger1987translational} 
in order to derive an integral equation for the diffusion of the 
polymer using the pre-averaging approximation, which was then 
manipulated to find a closed form equation for the diffusion 
in the long chain limit. This gave scaling laws for key parameters 
of the polymer at different scales. The developed multi-resolution
approach keeps the rms end-to-end distance and the diffusion of 
the polymer invariant to the choices of how we split the polymer 
up into different resolutions. These scaling laws have been then 
supported by illustrative simulations, which used an adapted 
version of the Ermak-McCammon algorithm~\citep{Ermak1978Brownian}.

This work has been looking at extensions to a polymer model in 
a theta solvent, which is not the most general state that a polymer 
can exist in. To extend this model further, it is of interest to 
include excluded volume forces to allow for the study of a good 
solvent, of which there has been much analytical work to derive 
terms for the rms end-to-end 
distance~\cite{freed1987renormalization,flory1949configuration} in 
the single-scale model. Another possible extension is to look at 
including additional forces between monomers to get more realistic 
spring forces which are used in many recent studies of
polymers~\cite{goldtzvik2016importance,schroeder2004effect,rosa2008structure}, 
for example to form a wormlike chain 
model~\cite{hagerman1988flexibility,marko1995stretching}.

There have also been many recent developments in the algorithms which 
are used to study polymers with hydrodynamic
interactions~\cite{miao2017iterative,fixman1986construction,ando2012krylov,geyer2009n}, 
which could improve the computational efficiency of the multi-resolution 
modelling even further than the Ermack-McCammon 
algorithm~\cite{Ermak1978Brownian}, which has been used here as 
a demonstration of the scalings.

\begin{acknowledgments}
\noindent
This work was supported by funding from the Engineering and 
Physical Sciences Research Council (EPSRC) [grant number EP/G03706X/1]. 
Radek Erban would also like to thank the Royal Society for a University
Research Fellowship.
\end{acknowledgments}

\end{document}